 \documentclass[prb,aps,twocolumn,showpacs]{revtex4} 
\usepackage{graphicx} 
\usepackage{tabularx}
\usepackage{dcolumn} 
\usepackage{color}
\usepackage{bm}
\usepackage{amsmath}
\usepackage{amssymb}

\begin{document} 


\title{Violation of the Pauli-Clogston limit in a heavy Fermion superconductor CeRh$_2$As$_2$
--Duality of itinerant and localized 4f electrons--
}

\author{Kazushige Machida} 
\affiliation{Department of Physics, Ritsumeikan University, 
Kusatsu 525-8577, Japan} 

\date{\today}

\begin{abstract}
We theoretically propose a mechanism to understand the violation  of the Pauli-Clogston limit for the
upper critical field H$_{\rm c2}$ observed in the Ce bearing heavy Fermion material CeRh$_2$As$_2$ from the view point of
spin singlet pairing. It is based on a duality concept, the dual simultaneous aspects of an electron: the itinerant part and localized
part of quasi-particles (QPs) originated from the 4f electrons of the Ce atoms. While the itinerant QPs directly participate
in forming the Cooper pairs, the localized QPs exert the internal field so as to oppose the applied field
through the antiferromagnetic exchange interaction between them. This is inherent in the dense Kondo
lattice system in general. We argue that this mechanism can be applied not only to
the locally noncentrosymmetric material CeRh$_2$As$_2$,  but also to globally inversion symmetry
broken Ce-based materials such as CePt$_3$Si. Moreover,  we point out that it also works for strongly Pauli limit violated 
spin triplet pairing systems, such as UTe$_2$.
 \end{abstract}

 
 
\maketitle 

\section{Introduction}
The newly found heavy Fermion superconductor (SC) CeRh$_2$As$_2$ is 
attracting enormous attention both experimentally~\cite{khim,hafner,hassinger,onishi,kimura0} 
and theoretically~\cite{new1,new2,new3,new4,new5,new6,new7,new8}
 because, compared with the SC transition temperature
T$_{\rm c}$=0.35K
both of the upper critical fields H$_{\rm c2}^{\rm c}\sim16$T for the $c$-axis and 
H$_{\rm c2}^{\rm ab}\sim2$T for the $ab$-plane exceed the Pauli-Clogston
limit estimated by the weak coupling BCS formula H$_{\rm P}^{\rm BCS}$
=1.84T$_{\rm c}\sim 0.6$T in the tetragonal crystal symmetry.
The degree of the violation of the Pauli limit H$_{\rm c2}^{\rm c}$/H$_{\rm P}^{\rm BCS}$$\sim$27 is
extraordinary.

Given that the local symmetry on the Ce sites breaks the inversion symmetry,
it is argued that  the spin singlet-triplet  mixing scenario to overcome the Pauli limitation
is realized in this compound~\cite{new1,new2,new3,new4,new5,new6,new7,
new8}. This scenario is an extended version 
designed for globally non-centrosymmetric SC materials~\cite{sigrist,sigrist2,sigrist3,sigrist4,sigrist5,sigrist6,sigrist7},
in particular on Ce heavy Fermion SC~\cite{chrisRMP} 
such as CePt$_3$Si~\cite{bauer,takeuchi,metoki,yogi}, CeIrSi$_3$~\cite{mukuda,settai}, 
CeRhSi$_3$~\cite{kimura}, and CeCoGe$_3$~\cite{settai2}. They also 
break the Pauli limitation and CeIrSi$_3$ exhibits a record high  H$_{\rm c2}$$\sim$45T
with  T$_{\rm c}$=2K under pressure~\cite{kimura}.
However, no firm experimental evidence has proven those theories so far.

There have been several known mechanisms to explain the Pauli limit violation
apart from the spin triplet pairing.
For example, the Fulde-Ferrell-Larkin-Ovchinnikov (FFLO) state can raise the Pauili limit,
but it is only within a factor of 2 or so of H$_{\rm c2}^{\rm c}$/H$_{\rm P}^{\rm BCS}$ at most~\cite{fflo}. 
An especially designed thin film system
with few layers shows the enhanced H$_{\rm c2}$ due to strong spin-orbit coupling,
leading to the so-called Ising superconductivity~\cite{gated,iwasa}, or twisted magic angle graphene
exhibits also the Pauli limit violation~\cite{magic}. 
Apparently, those are not appropriate for the present three dimensional bulk systems.

To understand the strong Pauli limit violation in CeRh$_2$As$_2$,
the following should be noted:

 \noindent
 (1) The phase diagram in $H$($\parallel$ c) vs $T$ is subdivided into
 the SC1 and SC2 phases~\cite{khim}  for low and high fields  separated by a first order line at $H$=4T
 as shown schematically in Fig.~1(a).
 The SC2 phase reaches 16T far beyond the Pauli limit with a large margin as mentioned.
 
 \noindent
 (2) Below  T$_{\rm c}$=0.35K, the antiferromagnetic order (AF) develops at 
 T$_{\rm N}$=0.25K whose detailed AF structure has not been determined yet~\cite{kibune,kitagawa}.
 This order disappears above the field $H^c>$4T applied along the $c$-axis~\cite{ogata},
 whose value approximately coincides with the SC1 and SC2 boundary line.
 
\noindent
 (3) By tilting the field direction from the $c$-axis towards the $ab$ plane
 by the angle $\theta$, the enhanced H$_{\rm c2}(\theta)$ quickly diminishes
 up to $\theta\sim 30^{\circ}$ beyond which  H$_{\rm c2}(\theta)$ smoothly tends to
  H$_{\rm c2}^{ab}$=2T for the $ab$-plane~\cite{hassinger}. 
Thus, the low field phase of the SC1, starting below 4T for the $c$-axis, is continuously 
connected to H$_{\rm c2}^{ab}$.

\noindent
 (4) According to the recent Knight shift (KS) experiments of $^{75}$As-NMR
 by the Ishida group~\cite{kibune,kitagawa,ogata}, not only for the SC1 phase, 
 but also  for the SC2 phase for the 
 $c$-axis, KS  does decrease below T$_{\rm c}$, negating a spin triplet phase.
 Note that KS also decreases for the $ab$-plane field~\cite{ishida}. 
 There is evidence neither for the
 spin-triplet pairing, nor the singlet-triplet mixing associated with local inversion symmetry
 breaking~\cite{new1,new2,new3,new4,new5,new6,new7,
new8}. This urges us to consider the Pauli limit violation within the spin-singlet
 framework, or more broadly a framework applicable to both singlet and triplet pairings.

\noindent
 (5) It is noteworthy to remind of the fact that LaRh$_2$As$_2$~\cite{landaeta} 
 with identical  locally non-centrosymmetric crystal structure
 and similar T$_{\rm c}\sim$0.3K, shows neither the enhanced H$_{\rm c2}$ 
 (H$_{\rm c2}^c$=10mT and H$_{\rm c2}^{ab}$=12mT),
nor the multiple phase diagram.
This means that the 4f electrons of the 
Ce atoms play crucial roles in those intriguing phenomena, in particular the
strong Pauli limit violation of H$_{\rm c2}^{\rm c}=16$T.

\noindent
 (6) Substantial magnetic moments are progressively induced with increasing applied fields
 at low $T$, i.e., $M_{\rm c}(M_{\rm ab})$=0.2 (0.4) $\mu_{\rm B}$/Ce for 
the $c$ ($ab$)-axis under $H$=15T.
In view of the dual nature of the 4f electrons of the Ce atoms in the present dense Kondo lattice 
material with T$_{\rm Kondo}\sim$30K,  
a part of the 4f electrons is localized to form the AF order and the other part is itinerant
to form a coherent Fermion state with heavy quasi-particle mass; 
It has a huge Sommerfeld coefficient $\gamma_{\rm N}\sim$1J/mol$\cdot$K$^2$.
The latter directly participates in the Cooper pair formation.
This duality or dichotomy of the 4f electrons is essential in unveiling the physics
 of CeRh$_2$As$_2$.
 
 All figures are schematic throughout the paper, intending to no quantitative meaning.

\section{Basic idea and assumptions}

To overcome the fundamental and seemingly unavoidable H$_{\rm c2}$ limitation due 
to the Pauli paramagnetic effect associated with a spin singlet pairing, we consider the effects of the 
localized moment $M(H)$ originating from the 4f electrons on the Ce atomic sites. 
This is to exert the internal field $J_{\rm cf}M(H)$ to the conduction
electrons through the c-f exchange interaction $J_{\rm cf}$ coming from the periodic Kondo
lattice Hamiltonian necessary for describing the heavy Fermion systems in general.
In the past, this interaction was considered to play several important and crucial roles 
in the coexistence  problems of magnetism and superconductivity. In the ferromagnetic case, 
it stabilizes the FFLO state
via the ferromagnetic molecular field~\cite{nakanishi} whereas in the
 AF case, it yields the suppressed H$_{\rm c2}$
below T$_{\rm N}$~\cite{nokura,matsubara,kato}.
This idea is somewhat similar to the Jaccarino-Peter mechanism~\cite{jaccarino}.

The sign of $J_{\rm cf}>$ 0 is generically
antiferromagnetic for our dense Kondo lattice systems, i.e., CeRh$_2$As$_2$, to
realize the Kondo effect which ultimately leads to the heavy Fermion phenomenology.
Thus, the effective internal field $H_{\rm eff}$ felt by the conduction electrons is written as

\begin{equation}
H_{\rm eff}(H)=H-J_{\rm cf}M(H),
\label{eff}
\end{equation}

\noindent
with $H$ being the applied external field.
We assume that in the AF order, the sublattice moment is parallel to the $c$-axis
although we know that the system is a magnetically easy $ab$ plane  XY type~\cite{kitagawa}.
 This conflicting situation sometimes 
happens in other Ce-Kondo materials~\cite{araki}.
Under the field parallel to the $c$-axis, via a first order transition, the AF 
flips the moment $M_0$ towards the $c$-axis at $H_{\rm FL}$ in general.
We assume $H_{\rm FL}$=4T, coinciding with
the field above in which the NMR experiment detects no AF.
Until this spin flop transition $H<H_{\rm FL}$=4T the total moment $M(H)$=0 in the normal state.
The magnetization process along the $c$-axis is schematically depicted in Fig.~1(b) where 
at $H_{\rm FL}$=4T, the moment jumps by $M_0$.
Thus, for the $c$-axis,

\begin{eqnarray}
M_c(H)&=&0 \qquad \qquad \qquad \qquad \quad\quad  \;{\rm for} \> H<H_{\rm FL} \nonumber\\
        &=&M_0+\chi_cH+\chi_c^{(3)}H^3 \quad \quad {\rm for}\> H\geqq H_{\rm FL}
\label{Mc}
\end{eqnarray}

\noindent
while for the $ab$-axis,

\begin{eqnarray}
M_{ab}(H)=\chi_{ab}H+\chi_{ab}^{(3)}H^3+\cdots,
\label{Ma}
\end{eqnarray}

\noindent
where $\chi_{i}$ and $\chi_{i}^{(3)}$ ($i=c$ and $ab$) are the linear and non-linear 
magnetic susceptibilities respectively.
By substituting $M(H)$ into Eq.~(\ref{eff}), we obtain

\begin{eqnarray}
H^c_{\rm eff}&=&H \qquad \qquad \qquad \qquad \quad\quad \quad\quad {\rm for} \> H<H_{\rm FL} \label{effc}\\
        &=&(1-\chi_cJ^c_{\rm cf})H-J^c_{\rm cf}(M_0+\chi_c^{(3)}H^3)  \;{\rm for}\> H\geqq H_{\rm FL}. \nonumber
\end{eqnarray}


\noindent
For $H\parallel ab$,

\begin{eqnarray}
H^{ab}_{\rm eff}(H)=(1-\chi_{ab}J^{ab}_{\rm cf})H-J^{ab}_{\rm cf}\chi_{ab}^{(3)}H^3+\cdots.
\label{effab}
\end{eqnarray}

\noindent
The cf-exchange interaction constants are anisotropic, i.e., $J^c_{\rm cf}\neq J^{ab}_{\rm cf}$ in general.
The following can be clearly observed: 

\noindent
(1) The external field is scaled by a factor $1-\chi J$ as expressed in Eqs.~(\ref{effc}) and (\ref{effab}).

\noindent
(2) The external field is reduced (enhanced) by a factor $JM_0$ for the antiferromagnetic $J_{\rm cf}>0$ 
(ferromagnetic $J_{\rm cf}<0$)
cf-coupling case, as expressed in Eq.~(\ref{effc}).

\section{GL theory}
\subsection{H$\parallel$c}

To see the effects of the scaling factor and the reduction for 
the external field on $H_{\rm c2}$, we employ the Ginzburg-Landau(GL)
theory given by

\begin{equation}
H_{\rm c2}(T)=\alpha_0\cdot (T_{\rm c0}-T),
\label{gl}
\end{equation}

\noindent
where the GL coefficient $\alpha_0(>$0) related to the effective mass determines the
slope of $H_{\rm c2}(T)$ at $T_{\rm c0}$.
With the effective field $H_{\rm eff}$ in place of $H$ in Eq.~(\ref{gl}), 
$H_{\rm eff,\rm c2}(T)=\alpha_0\cdot (T_{\rm c0}-T)$ is obtained.
After plugging Eq.~(\ref{effc}) into it, we find
for $H\parallel c$

\begin{eqnarray}
H_{\rm c2}^c(T)&=&\alpha_0^c\cdot(T_{\rm c0}-T) \qquad \quad {\rm for} \> 0<H<H_{\rm FL} \nonumber \\
                    &=&{\alpha_0^{c}\over {1-\chi_cJ^c_{\rm cf}}}\cdot(T_{\rm c}-T) \quad{\rm for}\> H\geqq H_{\rm FL}
\label{hc2c}
\end{eqnarray}

\noindent
with

\begin{eqnarray}
T_{\rm c}=T_{\rm c0}+{J^c_{\rm cf}\over \alpha_0^c}M_0.
\label{tc}
\end{eqnarray}

\noindent
Two factors raise $H_{\rm c2}^c(T)$, one through the effective mass and
the other through $T_{\rm c}$.
From now on we neglect the non-linear susceptibility $\chi^{(3)}$ term for simplicity.
For $H\parallel$$ab$,  we find

\begin{eqnarray}
H_{\rm c2}^{ab}(T)={\alpha_0^{ab}\over {1-\chi_{ab}J^{ab}_{\rm cf}}}\cdot(T_{\rm c0}-T).
\label{hc2ab}
\end{eqnarray}

We show a schematic $H_{\rm c2}^c(T)$ in Fig. 1(a).
As can be observed from this, when $H_{\rm c2}^c(T)$ started from $T_{\rm c0}$
with the slope $dH_{\rm c2}^c(T)/dT=-\alpha_0^c$ reaches $H=H_{\rm FL}$,
it jumps by $J^c_{\rm cf}M_0$ which is estimated by $\sim$4T later. 
Then, according to Eq.~(\ref{hc2c}),
$H_{\rm c2}^c(T)$ is enhanced by the scaling factor, namely

\begin{eqnarray}
H_{\rm c2}^c(T=0)={\alpha_0^{c}T_{\rm c}\over {1-\chi_cJ^c_{\rm cf}}},
\label{hc2c0}
\end{eqnarray}

\noindent
with the enhanced slope

\begin{eqnarray}
{dH_{\rm c2}^c\over dT}=-{\alpha_0^{c}\over {1-\chi_cJ^c_{\rm cf}}}.
\label{hc2c0'}
\end{eqnarray}

\noindent
Notice that the high field part of $H_{\rm c2}^c(T)$ has the enhanced $T_{\rm c}$
given in Eq.~(\ref{tc}). Those factors compound to push $H_{\rm c2}$ to a higher field.

\begin{figure}
\includegraphics[width=8cm]{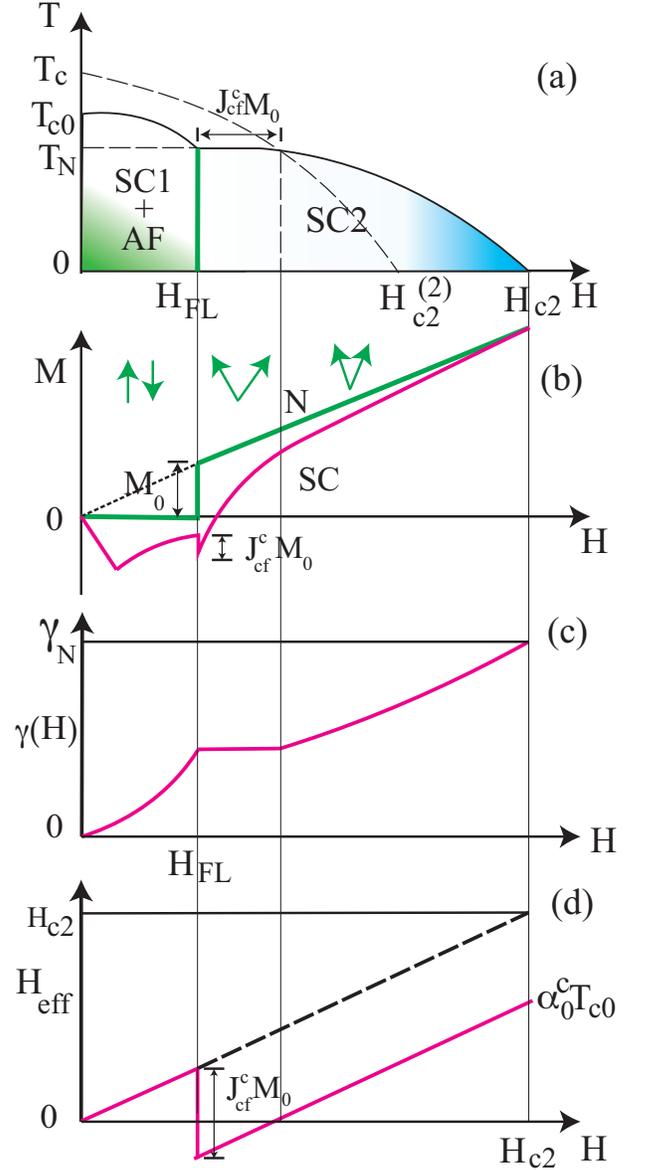}
\caption{The field dependences of various quantities for $H$$\parallel$c-axis.
(a) H$^c_{\rm c2}$ vs T phase diagram. SC1 starts at $T_{\rm c0}$ with the slope $dH_{\rm c2}/dT=
-\alpha_0^c$ and reaches $\alpha_0^cT_{\rm c0}$ at $T$=0.
$H_{\rm c2}^{(2)}=\alpha_0^c(T_{\rm c}-T)$ for SC2. $H_{\rm c2}$ ultimately reaches 
$\alpha_0^cT_{\rm c}/(1-J^{\rm c}_{\rm cf}\chi_c)$ at $T=0$ with the enhanced slope
-$\alpha_0^c/(1-J^{\rm c}_{\rm cf}\chi_c)$. Note a jump by $J^{\rm c}_{\rm cf}M_0$.
(b) Magnetization processes for the normal (N) and SC states.
In the normal state $M$=0 for $H<H_{\rm FL}$ and jumps by $M_0$ at 
$H_{\rm FL}$ via the first order spin flop transition.
In the SC it exhibits the negative jump by -$J^{\rm c}_{\rm cf}M_0$ on top of SC diamagnetic
background. Here we sketch the AF spin configurations for each field region where
at $H=0$ the moment points to the $c$ direction. 
(c) Field dependence of $\gamma(H)$. In SC1 for 0$<$$H$$<$$H_{\rm FL}$
it shows a strong Pauli affected curve with a concave curvature~\cite{machida}.
Corresponding to the $H_{\rm c2}$ jump,  $\gamma(H)$ stays a constant and
then gradually increases up to the normal value $\gamma_{\rm N}$ at $H^c_{\rm c2}$.
(d) The effective field $H_{\rm eff}(H)=H$ for 0$<$$H$$<$$H_{\rm FL}$.
After showing the negative jump  by -$J_{\rm cf}M_0$, $H_{\rm eff}(H)$
grows linearly in $H$ and reaches $\alpha^c_0T_{\rm c0}$ at $H^c_{\rm c2}$
far below the un-enhanced case drown by the dashed line.
}
\label{f1}
\end{figure}

There is no distinction between the SC1 phase for $0<H<H_{\rm FL}$ and the SC2 phase
for $H>H_{\rm FL}$ in the pairing symmetry in our scenario. Note, however, 
that the SC1 phase coexists with AF below T$_{\rm N}$. 
Various observed thermodynamic anomalies~\cite{khim} at $H$=4T
such as ac-susceptibility $\chi_{ac}(H)$, $M_c(H)$, and magnetostriction are due to the first order
phase transition associated with the AF spin flop transition $H_{\rm FL}$
although it was interpreted as the pairing symmetry change from a spin singlet to triplet
pairing~\cite{khim,hafner,hassinger,landaeta}.

In Fig.~1(b) we illustrate the magnetization curves both for the SC and normal states.
At $H=H_{\rm FL}$ vis the first order spin flop transition $M_c(H)$ exhibits
a jump by $M_0$ in the normal state. Correspondingly, in the SC state
a negative jump by $-J^c_{\rm cf}M_0$ appears.
According to the data~\cite{khim}, the magnetization curve exhibits a kink-like anomaly at $H$=4T
in the superconducting state. We interpret it as a first order negative jump. 

As shown later, the SC1 phase is strongly suppressed
by the Pauli paramagnetic effect characterized by a large Maki parameter,
compared with the SC2 phase. The
Sommerfeld coefficient $\gamma(H)$ exhibits a characteristic 
downward curvature~\cite{machida0,machida}
up to $H<H_{\rm FL}$ as displayed in Fig.~1(c).
This is followed by a plateau corresponding to the $H_{\rm c2}^c$ jump
above which $\gamma(H)$ grows slowly and monotonically.
The existing data~\cite{onishi} for $\gamma(H)$ and thermal conductivity $\kappa(H)$
at the lowest temperature limit both exhibit a similar behavior.
Those data are consistent with the above picture.

In Fig.~1(d) we summarize the field evolution of effective field $H_{\rm eff}(H)$;
For $H<H_{\rm FL}$, $H_{\rm eff}(H)=H$. Then after showing the negative jump 
of $-J^c_{\rm cf}M_0$, it grows linearly up to $H^c_{\rm c2}$ where 
$H_{\rm eff}=\alpha_0^cT_{\rm c0}$. This value is far less than the reached $H^c_{\rm c2}(T=0)$
given by Eq. (\ref{hc2c0}).

\subsection{H$\parallel$ab}

Let us consider the case of $H$$\parallel$$ab$ whose direction is perpendicular to the
AF moment. In this case $M_{ab}(H)=\chi_{ab}H$ because the sublattice moment continuously
rotates towards the field direction. As illustrated in Fig.~2(a), $H^{ab}_{\rm c2}$  
given by Eq.~(\ref{hc2ab}) is enhanced by the factor $1-\chi_{ab}J^{ab}_{\rm cf}$.
This is compared with the corresponding orbital limit value 
$H^{ab}_{\rm c2}(orb)=\alpha_0^{ab}T_{\rm c0}$.
This means that even in the paramagnetic state under suitable conditions,
the violation of the Pauli limit is possible, implying that the present violation mechanism is 
quite generic applicable to other systems. In Fig.2 we summarize 
the corresponding behaviors for this orientation,
and in Fig. 2(c) we schematically plot $\gamma(H)$ with the Maki parameter 
$\mu_{\rm M}$=0.8~\cite{machida}.
Note that the effective field $H_{\rm eff}(H^{ab}_{\rm c2})=\alpha_0^{ab}T_{\rm c0}$ is reduced
by the factor $1-\chi_{ab}J^{ab}_{\rm cf}$ as depicted in Fig. 2(d).

\begin{figure}
\includegraphics[width=6.5cm]{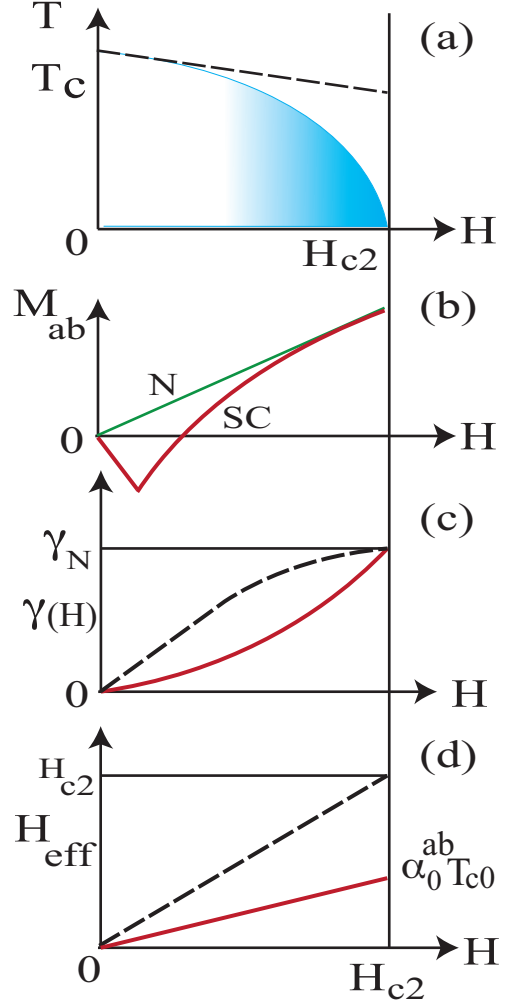}
\caption{The field dependences of various quantities for $H$$\parallel$ab-plane.
(a) H$^{ab}_{\rm c2}$ vs T phase diagram
where it is enhanced by $H^{ab}_{\rm c2}(orb)/{(1-\chi_{ab}J_{\rm cf})}$
from the orbital $H^{ab}_{\rm c2}(orb)$. The dashed line denotes the initial slope.
$H^{ab}_{\rm c2}(T=0)$ is low because of the paramagnetic effect.
(b) Magnetization processes for the normal and SC states.
In the normal state $M_{ab}(H)=\chi_{ab}H$.
In the SC $M_{ab}(H)$ consists of the superconducting diamagnetic contribution and the paramagnetic contribution due to the localized moments.
(c) Field dependence of $\gamma(H)$.
it shows a strong Pauli affected curve with a concave curvature with $\mu_{\rm M}$=0.8~\cite{machida}.
The dashed curve indicates $\gamma(H)$ for the s-wave case with a full gap without the
Pauli paramagnetic effect $\mu_{\rm M}=0$~\cite{machida}.
(d) The effective field $H_{\rm eff}(H)=H$ grows linearly 
in $H$ and reaches $\alpha^{ab}_0T_{\rm c0}$ at $H^{ab}_{\rm c2}$
far below the un-enhanced case drown by the dashed line.}
\label{f2}
\end{figure}

\subsection{Field tilting from the $c$-axis to the $ab$ plane}

When tilting the field direction from the $c$-axis to the $ab$-plane by $\theta$, $H_{\rm c2}(\theta)$
decreases quickly from $H_{\rm c2}^c$=16T to $H_{\rm c2}(\theta=30^{\circ})=4$T~\cite{hassinger}.
This finding is analyzed within the present framework. This can be attributed to 
the angle dependence of the magnetization jump  $M_0(\theta)$ at $H_{\rm FL}$
as shown in the inset of Fig.3.
Namely, $H_{\rm c2}(\theta)$ is evaluated near the small angle $\theta$ as

\begin{eqnarray}
H_{\rm c2}(\theta)={\alpha_0^{c}\over {1-\chi_cJ^c_{\rm cf}}}\cdot(T_{\rm c}(\theta)-T),
\label{hc2theta}
\end{eqnarray}

\noindent
with $T_{\rm c}(\theta)=T_{\rm c0}+{J^c_{\rm cf}\over \alpha_0^c}M_0(\theta)$
for $H\geqq H_{\rm FL}$.  This reduces to Eq.~(\ref{hc2c}) when $\theta$=0 for the $c$-axis.
 As seen below, $H_{\rm FL}$ hardly changes with $\theta$
according to the standard phenomenological theory for the spin flop transition~\cite{kanamori}.
As will be explained, the AF is quite fragile for the tilted field because the competing two anisotropies; 
$K_{\rm AF}(>0)$ aligns the sublattice moment along the $c$-axis
and $K$ is the intrinsic anisotropy reflecting the fact 
that $\chi_{ab}=2\chi_{c}$ in the paramagnetic state~\cite{khim}.
This is characterized by an easy plane XY anisotropy~\cite{kitagawa}.
The spin flop transition is estimated by comparing the two free energies $f_c$ and $f_{ab}$
for the AF state  with the moment along the $c$ and $ab$ directions, respectively.
Under the tilted field $\theta$, those are given by

\begin{eqnarray}
F_c&=&-{1\over 2}\chi_{ab}\sin^2\theta\cdot H^2-K_{\rm AF},\nonumber\\
F_{ab}&=&-{1\over 2}\chi_{c}\cos^2\theta\cdot H^2-K.
\label{free}
\end{eqnarray}

\noindent
By equalizing the two energies, we obtain

\begin{eqnarray}
H_{\rm FL}=\sqrt{{2(K_{\rm AF}-K)\over {\chi_c\cos^2\theta-\chi_{ab}\sin^2\theta}}}.
\label{FL}
\end{eqnarray}

\noindent
This reduces to the standard expression~\cite{kanamori} of $H_{\rm FL}=\sqrt{{2(K_{\rm AF}-K)\over {\chi_c}}}$
when $\theta$=0.
Equation (\ref{FL}) indicates an absolute  instability of the AF with the moment along the $c$-axis.
This analysis is only meaningful for $\chi_c\cos^2\theta-\chi_{ab}\sin^2\theta>0$,
namely, 

\begin{eqnarray}
\theta_{\rm cr}\leqq \tan^{-1}\sqrt{\chi_c/\chi_{ab}}=\tan^{-1}\sqrt{1/2}=35.2^{\circ}.
\label{theta}
\end{eqnarray}

\noindent
Beyond $\theta_{\rm cr}$ the magnetic system may enter the paramagnetic state.
Thus it is conceivable that towards this critical angle the jump of the moment $M_0(\theta)$
decreases. According to our analysis of $H_{\rm c2}(\theta)$, we predict that it
decreases linearly in $\theta$ and vanishes around $\theta_{\rm cr}$,
as shown in the inset of Fig. 3. This can be verified experimentally.

It is noted from Fig.~3 that 
upon increasing $\theta$, (1) As $M_0(\theta)$ diminishes, the enhanced 
$H_{\rm c2}(\theta)$ quickly decreases because $T_{\rm c}(\theta)$ 
given by Eq.~(\ref{hc2theta}) drops.
(2) While $H_{\rm FL}$ is nearly independent of $\theta$, the first order transition
temperature $T_{\rm FL}(\theta)$ becomes lower because the orbital limit 
$H^{\rm orb}_{\rm c2}(\theta)=\alpha(\theta)\cdot(T_{\rm c0}-T)$ 
with the effective mass $\alpha(\theta)$
decreases from $H^{\rm orb}_{\rm c2}(\theta=0)$=4T to $H^{\rm orb}_{\rm c2}(\theta=90^{\circ})=2$T 
 according to the effective mass model discussed later.
 (3) Thus, above $\theta\sim30^{\circ}$, $H_{\rm c2}(\theta)$ cannot be enhanced
 simply because $H^{\rm orb}_{\rm c2}(\theta>30^{\circ})$ is less than $H_{\rm FL}$=4T,
 namely, it fails to reach the spin flop transition field.
 
 \begin{figure}
\includegraphics[width=6.5cm]{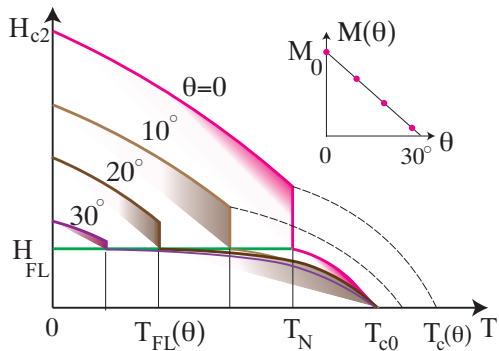}
\caption{
Angle dependences of $H_{\rm c2}(\theta)$ where $\theta$ is
the angle from the $c$-axis towards the $ab$-plane.
For $\theta=0$, $H^c_{\rm c2}$ starting from $T_{\rm c0}$
meets the spin flop transition line denoted by the green line,
and  it jump vertically around the point at ($T_{\rm N}$, $H_{\rm FL}$).
Then $H^c_{\rm c2}$ follows the dashed curve
with the enhanced $T_c(\theta)$=$T_{\rm c0}$+$
{J^c_{\rm cf}\over \alpha_0^c}M_0(\theta)$
and reaches the enhanced  $H^c_{\rm c2}(T=0)$ value given by Eq.~(\ref{hc2c0}).
Upon increasing 
$\theta$ because the initial slopes at $T_{\rm c0}$ decreases according to the
effective mass model, $T_{\rm FL}(\theta)$ is progressively lowering.
Beyond $\theta>30^{\circ}$ it fails to meet the $H_{\rm FL}$ line denoted by the green horizontal line.
Thus no enhanced $H_{\rm c2}$ occurs. The inset shows the predicted 
behavior of the magnetization jump $M_0$ as a function of $\theta$.
}
\label{f3}
\end{figure}

\subsection{Pauli paramagnetic effect and $J_{\rm cf}$ values}

The orbital limit $H^{\rm orb}_{\rm c2}$=17T and =8T for the $c$ and $ab$-axis
estimated from their initial slopes at $T_{\rm c0}$ are suppressed to
4T and 2T respectively~\cite{khim}. This is because of the Pauli paramagnetic effect
signified by the Maki parameter $\mu_{\rm M}$. 
This $\mu_{\rm M}$ is evaluated by employing an empirical formula derived
by the microscopic Eilenberger theory~\cite{machida} based on the effective 
mass model.

\begin{eqnarray}
H_{\rm c2}(\theta)={H^{\rm orb}_{\rm c2}(\theta=90^{\circ})\over 
{\sqrt{\Gamma^2\cos^2\theta+\sin^2\theta+2.4\mu^2_{\rm M}}}},
\label{mass}
\end{eqnarray}

\noindent
where $\Gamma$ is the effective mass anisotropy for the orbital limit $H^{\rm orb}_{\rm c2}$.
Substituting the above values for the $c$ and $ab$-axes, we determine $\Gamma$=1.75
and $\mu_{\rm M}$=2.5. This large Maki parameter gives rise to the first order transition
for ordinary superconductors.
Here because of the field scaling $H_{\rm eff}=(1-\chi J)H$, the effective Maki parameter
is reduced to $\mu=(1-\chi J)\mu_{\rm M}$ because 

\begin{eqnarray}
H_{\rm P}={H^{\rm BCS}_{\rm P}\over {1-\chi J}}.
\label{mass}
\end{eqnarray}

\noindent
For $H$$\parallel$$c$, 
$\mu_c$=0.4 and $(1-\chi_c J^c_{\rm cf})$=0.159, and for $H$$\parallel$$ab$,
$\mu_{ab}$=0.8 and $(1-\chi_{ab}J^{ab}_{\rm cf})$=0.32 with $H^{\rm BCS}_{\rm P}$=1.84$T_{\rm c0}$=0.64T.
Those moderate Maki parameter values avoid the first order transition at 
$H_{\rm c2}$ as observed.
We regard that their upper critical fields are both Pauli limited: $H^c_{\rm P}$=4.0T and
$H^{ab}_{\rm P}$=2.0T. Utilizing the observed susceptibilities~\cite{khim}  
$\chi_{c}$=0.016$\mu_{\rm B}$/T and $\chi_{ab}$=0.029$\mu_{\rm B}$/T,
 we obtain $J^c_{\rm cf}=52.5T$/$\mu_{\rm B}$ and $J^{ab}_{\rm cf}=23.4T$/$\mu_{\rm B}$.
 Their anisotropy $J^c_{\rm cf}/J^{ab}_{\rm cf}$=2.2.
This yields the $H^{c}_{\rm c2}$ jump: $J^c_{\rm cf}M_0$=52.5$\times\chi_cH_{\rm FL}$=3.6T
at $H^c=H_{\rm FL}$.

\section{Possible application to other materials}

Having performed the detailed analysis on CeRh$_2$As$_2$, we turn to other superconductors
that break the Pauli limit to apply the present scenario. As mentioned in Introduction, for the noncentrosymmetric 
Ce heavy Fermion superconductors~\cite{chrisRMP}, CePt$_3$Si, CeIrSi$_3$, 
CeRhSi$_3$ and CeCoGe$_3$ are possible
candidates because (1) our theory requires neither local and global inversion 
symmetry breaking in the crystalline structure. 
(2) Because those are all dense Kondo lattice systems,
the 4f electrons of the Ce atoms have the dual nature: itinerant and localized characters.
In fact, they also exhibit AF order above the superconducting transition, meaning that
the 4f electrons of the Ce atoms are localized. 
(3) The cf exchange coupling constants $J_{\rm cf}$ for those systems are expected to be antiferromagnetic,
thus the effective field is reduced from the applied  external field, enhancing the Pauli limit.
Those three conditions satisfy precisely the requirement for the violation of the Pauli limit
as explained above.

To facilitate future investigations further, we briefly examine CePt$_3$Si
with T$_{\rm c}$=0.75K.
As $H^{\rm BCS}_{\rm P}$=1.38T, the enhancement factor for the $c$-axis
$H^c_{\rm c2}(T$$=$$0)$/$H^{\rm BCS}_{\rm P}$=5T/1.38T=3.62, and thus
(1-$\chi_cJ^c_{\rm cf}$)=0.276. By knowing that 
$\chi_c$=0.025$\mu_{\rm B}$/T~\cite{takeuchi}, we find
$J^c_{\rm cf}$=29.0T/$\mu_{\rm B}$.
Similarly, for the $ab$-plane, the corresponding values are
$H^{ab}_{\rm c2}$($T$=0)=3T, and $\chi_{ab}$=0.02$\mu_{\rm B}$/T,
which yield $J^{ab}_{\rm cf}$=23.0T/$\mu_{\rm B}$.
The obtained cf exchange constants are similar numbers to those of 
CeRh$_2$As$_2$ as mentioned above,
suggesting that the same mechanism for the violation of the Pauli limit is working here.
The record high  H$_{\rm c2}$$\sim$45T
with  T$_{\rm c}$=2K under pressure in CeIrSi$_3$~\cite{kimura} may be within our reach
although we do not have further  experimental information for the detailed analysis.

It may be interesting to compare the present Kondo systems with the materials
where the obvious localized moments embedded in the conduction electrons
exert the field compensated internal field through the
Jaccarino-Peter  mechanism~\cite{jaccarino}.
For example, the Chevrel system Eu$_x$Sn$_{1-x}$Mo$_6$S$_8$~\cite{fischer} has the compensation
field -30T with the Eu localized moment, giving rise to $J_{\rm cf}$=8$\sim$9T/$\mu_{\rm B}$.
The exchange constant  J$_{\pi-d}$=2.3$\mu_{\rm B}$/T in an organic SC: $\kappa$-(BETS)$_2$FeBr$_4$
is estimated directly by NMR Knight shift experiment~\cite{takigawa}.
In this compound the field induced SC is observed around 15T with T$_{\rm c}$=0.3K. 
The present exchange constant $J_{\rm cf}$ is an order of magnitude
larger than those of non-Kondo materials.

We point out also the case where $J_{\rm cf}$ is ferromagnetic in TmNi$_2$B$_2$C~\cite{morten}.
According to the small angle neutron scattering (SANS) experiment~\cite{morten}, the
internal field differs from the applied field because the vortex lattice constant 
reflects directly the internal field, not applied field. 
Thus the measurement shows that the internal field is larger than the
applied field~\cite{morten}, indicating that the Tm localized moment 
{\it enhances} the applied field by $\sim$10$\%$, the opposite of the present CeRh$_2$As$_2$ case. 
The exchange constant is ferromagnetic.
It is understood that heavy Fermion SC is guarantied for $J_{\rm cf}$ to be antiferromagnetic
in general, satisfying one of the criteria for the violation of the Pauli limit.

\section{Conclusion and prospects}

As for CeRh$_2$As$_2$, it is desirable to perform experiments to better characterize the
phase boundary between SC1 and SC2 for $H$$\parallel$$c$ at 4T
because it was interpreted as a spin singlet-triplet pairing change~\cite{khim,hafner,hassinger,onishi}.
According to our theory, this is nothing but the spin flop transition via a first order.
The AF moment is assumed to point to the $c$-direction as a fundamental assumption in our theory. 
This can be verified by various methods, including neutron diffraction experiment.
As predicted in Fig. 3 the magnetization jump $M_0(\theta)$ at $H_{\rm FL}$=4T vanishes 
quickly by rotating the applied field from the $c$-axis towards the $ab$-plane up to $\theta$=30$^{\circ}$.
This is an important prediction to verify our scenario because the enhancement of $H_{\rm c2}(\theta)$
near the $c$-axis is closely correlated with $M_0(\theta)$.
Obviously, the gap structure should be characterized more precisely, either full gap or
nodal structure. There are several established spectroscopic methods to probe, such
as the field-angle dependent specific heat experiment~\cite{miranovic},
or the scanning tunneling spectroscopy to probe the local density of states~\cite{hayashi,ichioka}.

More generally, apart from CeRh$_2$As$_2$, the present theory on the violation mechanism
of the Pauli limit can be applied to other SC's,
in particular to the Ce containing Kondo systems~\cite{chrisRMP}, including 
CePt$_3$Si~\cite{bauer,takeuchi,metoki,yogi}, CeIrSi$_3$~\cite{mukuda,settai}, 
CeRhSi$_3$~\cite{kimura} and CeCoGe$_3$~\cite{settai2} as mentioned.
Here the duality of the 4f electrons of the Ce atoms is essential where the localized aspect 
produces the antiferromagnetic exchange field to cancel the applied field, and the
itinerant aspect produces the heavy Fermions. Both aspects are crucial to attain the
high field superconductivity beyond the Pauli-Clogston limit.
The extremely enhanced $H_{\rm c2}$ observed in those materials largely remains unexplained so far.
We propose several experiments on these superconductor to establish
the generality of our idea on the violation mechanism:
(1) To probe the actual internal field, or magnetic induction, which is a non-trivial task,
the Knight shift experiment of NMR is one of the direct methods.
In fact, it is applied successfully to probe the compensation field in the
organic superconductor~\cite{takigawa}.
(2) As mentioned above, the SANS experiment is also powerful to verify the internal field
because the vortex lattice spacing directly reflects the internal field via the flux quantization rule~\cite{morten}.

We should point out a common and unexpected feature between two singlet and triplet superconductors
where both are driven and reinforced by the incipient magnetization;
The $H_{\rm c2}$ enhancement in a spin singlet pairing here is analogous 
to the physics~\cite{machida2,machida3,machida4} in spin triplet pairing
in a series of magnetically polarized superconductors: UGe$_2$, URhGe, UCoGe, and
UTe$_2$, where the field reinforced $H_{\rm c2}$ is observed.
While the magnetization $M(H)$ is coupled through the exchange interaction $J_{\rm cf}$
in the form $M(H)J_{\rm cf}$
on the conduction electrons in a singlet case, it directly couples with a triplet pairing vectorial order
parameter $\vec\eta$ in the form of $\kappa {\vec M}(H)$$\cdot$$\vec\eta$$\times$$\vec\eta^{\star}$.
This common field-reinforced SC 
feature is deeply rooted in the duality nature of the f-electrons, itinerant and localized.

Finally, it should be noticed that the present mechanism of the Pauli-Clogston limit violation has been
applied  so far to the spin singlet pairing case in mind,  but it can work in the spin  triplet pairing as well
without any alternation.
Thus, it might be interesting to varify whether or not the observed extremely high H$_{\rm c2}$
enhancement over the Pauli limit; $60{\rm T}/1.84T_{\rm c}\sim 22$ in UTe$_2$ with T$_{\rm c}=1.5K$  needs this
mechanism in addition to the spin triplet pairing symmetry.

\begin{acknowledgements}
The author sincerely thanks K. Ishida and S. Kitagawa for enlightening discussions and
for sharing the data before publication, which were crucial for forming the present idea, and
T. Sakakibara and Y. Machida for their help in widening the scopes of the present theory.
The author is indebted to Editage for English language editing.
This work is supported by JSPS-KAKENHI, No. 17K05553 and No.21K03455.
\end{acknowledgements}

\end{document}